\documentclass[reprint,amsmath,amssymb,aps,prl,superscriptaddress, showpacs, eprint]{revtex4-2}
\usepackage[utf8]{inputenc}

\usepackage{amsmath,amssymb,ascmac,fancybox, multirow}
\usepackage{color}
\usepackage{graphicx}
\usepackage{url}
\usepackage{siunitx}
\usepackage{bm}
\usepackage{physics}
\usepackage{mathtools}
\usepackage{hyperref}
\usepackage[T1]{fontenc}
\usepackage{ulem}
\graphicspath{figures/}
\newcommand{\kB}{k_{\mathrm{B}}}

\newcommand{\ene}{\varepsilon}
\renewcommand{\vec}{\vb*}

\newcommand{\chiP}{\chi}

\begin{document}
\date{\today}
\title{Quantum weight}
\author{Yugo Onishi}
\affiliation{Department of Physics, Massachusetts Institute of Technology, Cambridge, MA 02139, USA}
\author{Liang Fu}
\affiliation{Department of Physics, Massachusetts Institute of Technology, Cambridge, MA 02139, USA}

\begin{abstract}
We introduce the concept of quantum weight as a fundamental property of insulating states of matter that is encoded in the ground-state static structure and measures quantum fluctuation in electrons' center of mass. 
We find a sum rule that directly relates quantum weight---a ground state property---with the negative-first moment of the optical conductivity above the gap frequency. Building on this connection to optical absorption, we derive both an upper bound and a lower bound on quantum weight in terms of electron density, dielectric constant, and energy gap. Therefore, quantum weight constitutes a key material parameter that can be experimentally determined from X-ray scattering.   
\end{abstract}

\maketitle

It is well known that quantum states of matter with an energy gap have vanishing dc longitudinal conductivity at zero temperature, while the optical conductivity is generally nonzero at frequencies above the gap. Interestingly, 
the ground state property of an insulating state still has important bearings on its optical conductivity. Consider the real part of longitudinal optical conductivity $\Re \sigma_{xx}(\omega)$, which determines the amount of optical absorption in the medium. It is well known that its zeroth moment known as the optical spectral weight is related to the electron density in the system through the $f$-sum rule~\cite{Kubo1957a}. 
Higher order moments of optical conductivity are much less studied. The negative-second moment is directly related to electric susceptibility through the Kramers-Kronig relation~\cite{onishi_universal_2024}.
Recently, we employed the negative-first moment of both longitudinal and Hall conductivities to derive a universal upper bound on the energy gap of (integer or fractional) Chern insulators~\cite{onishi_fundamental_2023}. 

While the Hall conductivity only exists in the presence of time reversal symmetry breaking, the longitudinal conductivity is present in all systems.  In a pioneering early work~\cite{souza_polarization_2000}, Souza, Willkins and Martin (SWM) showed that the negative-first moment of $\sigma_{xx}(\omega)$ is related to a basic property of quantum insulators that was expressed in terms of the quantum metric of many-body ground states over twisted boundary condition.  
This quantity was recently termed ``quantum weight'' because it connects quantum metric and optical weight \cite{onishi_fundamental_2023}. 
Despite its ubiquitous presence in solids, the quantum weight has not received adequate attention, and to our knowledge,  %
its value has not been experimentally determined for any material. %

In this work, we provide a new and general definition of quantum weight for all insulators as the quadratic coefficient of the ground-state static structure factor $S_{\vec{q}}$ at small $\vec{q}$. For periodic systems, $S_{\vec{q}}$ at any $\vec q$ other than reciprocal-lattice points vanishes in the classical limit ($\hbar =0$), and is nonzero
only because of quantum fluctuation in electron position. Therefore, the new definition of quantum weight manifests its purely quantum-mechanical origin, and allows it to be experimentally measured by X-ray scattering. 

Using fluctuation-dissipation theorem~\cite{callen_irreversibility_1951}, we show that as a ground state property, the quantum weight defined here is directly related to the negative-first moment of optical conductivity through the Planck constant. This establishes the equivalence between the old and new definition of quantum weight in terms of optical conductivity and static structure factor respectively.

Next, we derive lower and upper bounds on the quantum weight, in terms of the electron density, the energy gap, and the dielectric constant. These bounds apply to any insulating system, and represent a universal relation between ground state property (quantum weight), optical response, thermodynamic response (dielectric constant), and the energy gap of the system. Remarkably, our bounds can provide a good estimate of the quantum weight of real materials.

We start by considering generalized optical weights $W^i$ for insulating states \cite{onishi_fundamental_2023}. $W^i$ is defined as the negative $i$-th moment of the absorptive part of the optical conductivity $\sigma(\omega)$,  
\begin{align}
W_{\alpha \beta}^i \equiv \int_0^{\infty} \dd{\omega} \frac{\sigma^{\rm abs}_{\alpha\beta}(\omega)}{\omega^i},  
 \label{eq:generalized}
\end{align}
where $\sigma^{\rm abs}_{\alpha \beta} \equiv (\sigma_{\alpha \beta} + \sigma^*_{\beta \alpha})/2$ is composed of the real part of longitudinal optical conductivity and the imaginary part of optical Hall conductivity. For isotropic two-dimensional systems, $\sigma^{\rm abs}_{\alpha \beta} = \Re(\sigma_{xx}) \delta_{\alpha \beta} + i \Im(\sigma_{xy}) \epsilon_{\alpha \beta}$. 
Throughout this work, we consider gapped systems so that the absorptive part of optical conductivity vanishes at low frequency: 
\begin{align}
    \sigma^{\rm abs}(\omega) &= 0 \quad \text{for $\hbar\omega \le E_g$}. \label{eq:sigma_Eg}
\end{align}
This condition defines the optical gap $E_g$ which must be equal to or greater than the spectral gap $\Delta$.
We will consider three optical weights: $W^0, W^1$ and $W^2$, and focus on their real part, which is associated with optical longitudinal conductivity and is present in any solids. 

$\Re W^0$ and $\Re W^2$ are related to the charge density and the electric susceptibility, respectively: 
\begin{eqnarray}
    \Re W^0 &=& \pi ne^2/(2m), \label{fsum} \\
    \Re W^2 &=& (\pi/2)\chiP. \label{w2}
\end{eqnarray}
Eq.(\ref{fsum}) is the well-known $f$ sum rule~\cite{Kubo1957} that relates the full optical spectral weight to the electron density $n$ and mass $m$. 
Eq.(\ref{w2}) was recently derived from the relation between conductivity and polarizability using the Kramers-Kronig relation \cite{onishi_universal_2024}. Here, $\chiP$ describes the polarization induced by a static external electric field,  and is directly related to the dielectric constant $\chiP=\epsilon_0 (\epsilon-1)$ with $\epsilon_0$ the vacuum permittivity.

Our focus in this work is on the real part of the negative-first moment, $\Re W^1$. As we will show, the optical weight $\Re W^1$ is directly related to the quantum fluctuation in electron position, even though this quantity itself can be defined even for classical systems. To motivate the discussion, let us first consider a system of highly localized electrons arising from strong potential and interaction effects. %
In the classical limit ($\hbar = 0$), we can treat electrons as point charges and the ground state is obtained by minimizing the sum of the potential energy and the electron-electron interaction energy (which only depend on electron position): 
\begin{eqnarray}
H_c = \sum_i V(x_i)+ \sum_{i,j} U(x_i - x_j),
\end{eqnarray} with $V$ the potential and $U$ the two-body interaction. 
On the other hand, the kinetic energy 
\begin{eqnarray}
H_0=\sum_i \frac{p_i^2}{2m}, %
\end{eqnarray} 
leads to quantum fluctuation in electron position, which we treat below by quantizing electron's motion around the ground state configuration.    

We expand $H_c$ up to the second order in the displacement of the electrons from the ground state position, which yields a spring constant matrix $k$: $H_c=E_c + \sum_{i,j} k_{ij}\delta x_i \delta x_j/2$. Diagonalizing $k$, we obtain the normal modes $x'_{\alpha}=\sum_i c_{\alpha i} \delta x_i$ (with $\sum_i c_{\alpha i} c_{\beta i}=\delta _{\alpha\beta}$) and the spring constant $k_{\alpha}$: $H_c=\sum_\alpha k_{\alpha} {x'_{\alpha}}^2/2$. %
Correspondingly, we can rewrite the total kinetic energy $H_K$ with the momentum conjugate to the normal modes, $p'_{\alpha}=\sum_i c_{\alpha i} p_i$, as $\sum_{\alpha}{p'_\alpha}^2/(2m)$. Then we obtain a collection of independent harmonic oscillators, one for each normal mode, 
\begin{align}
    H &=H_0 + H_c \nonumber \\
    &\approx \sum_\alpha \frac{{p'_\alpha}^2}{2m} + \frac{1}{2}k_\alpha {x'_\alpha}^2 + \dots. ~\label{eq:Ham_strongly_localized}
\end{align}
The frequency for each mode is $\omega_\alpha=\sqrt{k_\alpha/m}$,  %
leading to the energy gap of $\hbar\omega_{\alpha}$ after quantization. 

For this system of strongly localized electrons, a uniform electric field couples to the center of mass displacement $\sum_i \delta x_i = \sum_\alpha Z_{\alpha} x'_{\alpha}$, and the real part of the optical conductivity is well known from that of the harmonic oscillator:  
\begin{align}
    \Re\sigma(\omega) &= \frac{1}{L}\sum_{\alpha}\frac{\pi Z_{\alpha}^2 e^2}{2m}\delta(\omega-\omega_\alpha),
\end{align}
with $Z_{\alpha}e=\sum_i c_{\alpha i}e$ the effective charge of the normal mode $\alpha$. Here, for simplicity we consider a one-dimensional system with length $L$ to illustrate the essential physics. 
It then follows 
\begin{align}
    \Re W^1 &= \frac{1}{L}\sum_{\alpha} \frac{\pi Z_{\alpha}^2 e^2}{2m\omega_\alpha}.
\end{align}
One can readily verify that $\sum_{\alpha} Z_{\alpha}^2/L = n$ with the electron density $n$, as expected from the $f$-sum rule. 

Interestingly, $\Re W^1$ is related through the Planck constant to the quantum fluctuation in the polarization of our system, which arises from electron's zero-point motion and is obtained by quantization: $p'_\alpha \rightarrow - i \hbar \partial'_\alpha$. Noting that the total polarization is given by 
$\delta P=e\sum_i \delta x_i = e\sum_\alpha Z_{\alpha}x_{\alpha}'$ and $\expval{x_{\alpha}' x_{\beta}'}=\delta_{\alpha\beta}\hbar/(2m\omega_\alpha)$, we can rewrite $\Re W^1$ as
\begin{align}
    \frac{\hbar}{\pi}\Re W^1 & =  \frac{1}{L} \expval{(\delta P)^2}, \label{eq:x2_W1} 
\end{align}
where $\expval{\dots}$ is the expectation value in the ground state.

This relation Eq.~\eqref{eq:x2_W1} is noteworthy for two reasons. The optical weight $\Re W^1$ in the left-hand side is finite even for classical systems, while the polarization fluctuation in the right-hand side is purely quantum-mechanical in nature and vanishes in $\hbar\to0$ limit. The left hand side involves optical conductivity at all frequencies, while the right hand side is a ground state property.  

In order to properly define polarization fluctuation for insulators in general, let us consider the static structure factor that measures equal-time density-density correlation in the ground state: $S_{\vec{q}} \equiv (1/V) (\expval{\hat{\rho}_{\vec{q}} \hat{\rho}_{-\vec{q}}} - \expval{\hat{\rho}_{\vec{q}}}\expval{\hat{\rho}_{-\vec{q}}})$.  
In particular, we focus on the quadratic coefficient of $S_{\vec q}$ at small nonzero $\vec q$:    
\begin{align}
    S_{\vec{q}} 
    &= \frac{e^2}{2\pi} K_{\alpha\beta}q_{\alpha}q_{\beta} + \dots, \label{eq:SW}
\end{align}
with $\hat{\rho}_{\vec{q}}=\int\dd{\vec{r}}e^{-i\vec{q}\vdot\vec{r}}\hat{\rho}(\vec{r})$ the charge density operator with wavevector $\vec{q}$ and  $V$ the volume of the system. We call the quadratic coefficient $K$ defined by Eq.~\eqref{eq:SW} \textit{quantum weight}.   
Importantly, $K$ can be nonzero only because of quantum fluctuation. In the classical limit ($\hbar =0$), the ground state of a periodic system is a periodic array of electron point charges, and correspondingly, the static structure factor is composed of $\delta$ functions centered at reciprocal lattice vectors and vanishes everywhere else, leading to $K=0$. 
Hence quantum weight measures the degree of ``quantumness'' in the insulating ground state. 

To make the connection between polarization fluctuation and quantum weight, note that polarization is related to charge density through $\nabla \cdot P = -\rho$. Therefore, 
one may identify $-iq\delta P = \rho_{q}$ and then quantum weight defined by Eq.~\eqref{eq:SW} represents the polarization (or center-of-mass position) fluctuation, in the ground state: $K = \expval{(\delta P)^2}/(2\pi e^2 V)$. In contrast to our treatment based on static structure factor, previous works treated polarization fluctuation in terms of the second cumulant moment of the position operator for systems with open or twisted boundary conditions~\cite{souza_polarization_2000, resta_polarization_2006}. %

Motivated by the case of strongly localized electron systems discussed above, we now establish a general relation between the optical weight $\Re W^1$ and the quantum weight $K$ encoded in ground-state static structure factor $S_{\vec{q}}$, which captures quantum fluctuation in electrons' center of mass of all insulators. 
Let us consider the response of an insulator to a time-dependent periodic potential $V_{\rm ext}$ with wavevector $\vec{q}$ and frequency $\omega$. The induced change in the density and the current response are characterized by the density-density response function $\Pi(\vec{q},\omega)$ and the conductivity tensor $\sigma(\vec{q},\omega)$ respectively: 
\begin{align}
    \rho(\vec{q},\omega) &= -\Pi(\vec{q},\omega) V_{\rm ext}(\vec{q},\omega), \\
    \vec{j}(\vec{q},\omega) &= \sigma(\vec{q},\omega) \vec{E}(\vec{q},\omega),
\end{align}
where $\vec{E}(\vec{q},\omega) = -i\vec{q}V_{\rm ext}(\vec{q},\omega)$ is the external electric field.
Due to the continuity equation 
 $ \pdv*{\rho}{t} + \nabla \cdot \vec{j} = 0$, 
$\sigma$ and $\Pi$ are directly related:  
\begin{align}
    \Pi(\vec{q},\omega) &= i\frac{q_\alpha q_\beta \sigma_{\alpha\beta}(\vec{q},\omega)}{\omega}. \label{eq:chi-sigma}
\end{align}
It is important to note that there is no singularity at small $\omega$ and at small $\vec{q}$ in Eq.~\eqref{eq:chi-sigma} for insulators~\footnote{We note that the static structure factor for metallic systems at small $\vec{q}$ is dominated by $\abs{q}$-linear term, in contrast to insulators where the leading order term is quadratic. %
}. 
Correspondingly, the negative-first moment of the real part of optical conductivity at ${\vec q}=0$ is related to the imaginary part of density response $\Pi({\vec q}, \omega)$ at small ${\vec q}$ integrated over the frequency:  
\begin{align}
    \int_0^\infty\dd\omega \Im\Pi(\vec{q},\omega) &= q_\alpha q_\beta \Re (W^1_{\alpha \beta}) + \dots. \label{im-chi}
\end{align}
By the fluctuation-dissipation theorem~\cite{callen_irreversibility_1951}, $\Im\Pi(\vec{q},\omega)$ is directly related to the dynamical structure factor: \begin{eqnarray}
    \Im \Pi(\vec{q}, \omega)=S(\vec{q}, \omega)/(2\hbar), \label{eq:Pi-S}
\end{eqnarray} 
where the dynamical structural factor is defined as 
  $  S(\vec{q},\omega) = \frac{1}{V}\int_{-\infty}^{\infty}\dd{t} e^{i\omega t} \expval{\hat{\rho}_{\vec{q}}(t)\hat{\rho}_{-\vec{q}}(0)}$. 
Here, $\hat{\rho}_{\vec{q}}(t)=\int\dd{\vec{r}}e^{-i\vec{q}\vdot\vec{r}}\hat{\rho}(\vec{r},t)$ is the density operator with wavevector $\vec{q}$ in the Heisenberg picture and $V$ is the volume of the system.

Combining Eq.~\eqref{eq:chi-sigma}, \eqref{im-chi} and \eqref{eq:Pi-S},
we obtain a general relation between optical conductivity and ground state static structural factor:
\begin{align}
     q_\alpha q_\beta \int_0^\infty \dd{\omega} \frac{\Re\sigma^{\rm abs}_{\alpha\beta}(\vec{q},\omega)}{\omega} 
    & = \frac{1}{2\hbar}\int_{-\infty}^\infty \dd{\omega} S(\vec{q}, \omega) \nonumber \\
     &%
     =\frac{\pi}{\hbar}S_{\vec{q}}
     \label{sum}
\end{align}
where we have used $S(\vec{q},\omega)=0$ for $\omega<0$ at zero temperature (for more details see Supplemental Materials).
Taking $\vec{q}\to0$ limit, we obtain a relation between optical weight $\Re W^1$ at $\vec q=0$ and the quantum weight defined above as the quadratic coefficient of $S_{\vec{q}}$:
\begin{align}
   \Re W^1 = \int_0^{\infty} \dd{\omega} \frac{\Re\sigma^{\rm abs}(\omega)}{\omega} = \frac{e^2}{2\hbar} K. \label{K-w1}
\end{align}
Eq.~\eqref{sum}, together with its $\vec q\rightarrow 0$ limit Eq.~\eqref{K-w1}, is a key result of our work. It constitutes a new optical sum rule relating the negative first moment of longitudinal optical conductivity to ground-state structure factor through the Planck constant, which is a generalization of the $f$ sum rule relating the optical spectral weight to the electron density.

The static structure factor $S_{\vec{q}}$ can be experimentally obtained from X-ray scattering experiments, while the optical weight $\Re W^1$ can be determined from the experimentally measured optical conductivity. The ratio between $S_{\vec{q}}/q^2$ and $\Re W^1$ yields a fundamental physical constant, the Planck constant. This provides a way of determining the Planck constant by optical spectroscopy measurements of basic material properties.

We have shown that quantum weight is both a fundamental ground state property of insulators and an important material parameter related to optical properties. We now further derive lower and upper bounds on quantum weight in terms of common material parameters: electron density, energy gap and dielectric constant. %
First, noting that the real part of the longitudinal optical conductivity is always non-negative and can be finite only for $\abs{\omega}\ge E_g/\hbar$, the following inequality among optical weights always holds: 
\begin{eqnarray}
 \Re W^{i+1}_{\alpha\alpha}\le \hbar \Re W^{i}_{\alpha\alpha}/E_g.     \label{wn_inequality} 
\end{eqnarray} 
Combining Eq.~\eqref{wn_inequality} with $i=0, 1$ and the expressions for optical weights $W^0, W^1, W^2$ shown in Eq.~\eqref{fsum},  \eqref{w2} and \eqref{K-w1}, we obtain the bound on the quantum weight: 
\begin{align}
\frac{\pi}{e^2} E_g\chi_{\alpha\alpha} \leq K_{\alpha\alpha}  \leq \frac{2\hbar^2}{e^2}\frac{W^0_{\alpha\alpha}}{E_g} \label{eq:K_ineq_W0}  . 
\end{align}
Here, $\alpha$ is the principal axis of the material.
The standard $f$-sum rule, $W^0=\pi ne^2/(2m)$ further leads to a universal bound on the quantum weight as 
\begin{align}
    \frac{\pi}{e^2} E_g\chi_{\alpha\alpha} \leq K_{\alpha\alpha}  \leq \frac{\pi n\hbar^2}{mE_g} \label{eq:K_ineq}  . 
\end{align}
This inequality should hold for any system regardless of the dimensionality of the system. Both the lower and upper bounds are saturated when optical conductivity is nonzero only at single frequency $\omega=\pm E_g/\hbar$, which we shall call single frequency absorption. We also note that the inequality \eqref{eq:K_ineq} remains valid even when the optical gap $E_g$ is replaced with the spectral gap, although the bounds would generally become less tight.

\begin{figure}
    \centering
    \includegraphics[width=0.9\columnwidth]{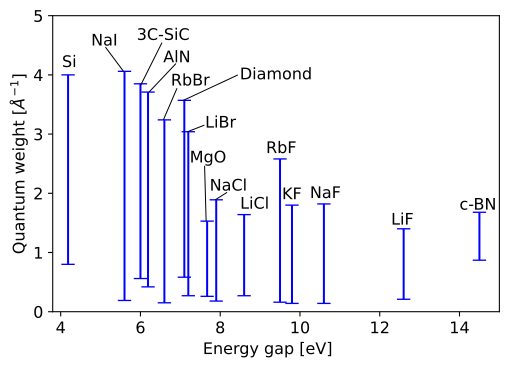}
    \caption{The bound on the quantum weight $K$ for real materials. As the energy gap, we used the gap obtained from optical measurements. The details of the used parameters are given in Ref.~\cite{onishi_universal_2024}.}
    \label{fig:Kbound}
\end{figure}

We calculated the bound on the quantum weight for real materials, and the results are shown in Fig.~\ref{fig:Kbound}. Our bound~\eqref{eq:K_ineq} gives a fairly good estimate of the quantum weight only from the electric susceptibility (or equivalently the dielectric constant), the energy gap, and the electron density. 
The most remarkable case is cubic boron nitride (c-BN). c-BN is an indirect gap insulator with direct gap \SI{14.5}{\electronvolt} and the dielectric constant $\epsilon=4.46$~\cite{madelung_semiconductors_2004}. From the dielectric constant, the electric susceptibility $\chi=\epsilon_0(\epsilon-1)$ is given by $\chi=\SI{3.1e-11}{\farad\per\metre}$. The cubic unit cell of c-BN has a lattice constant \SI{3.62}{\angstrom} and contains 4 boron and nitrogen atoms, each having 5 and 7 electrons, hence the total electron density is $n=\SI{1.02e24}{\per\cubic\centi\metre}$. Then the full spectral weight is given by $W^0_{aa}=\SI{4.5e22}{\per\metre\per\second\per\ohm}$. Since $K$ and $\chi$ are isotropic in c-BN, we find that the quantum weight $K_{aa}$ is bounded to a remarkably narrow interval: $\SI{0.9}{\per\angstrom}\le K_{aa}\le\SI{1.7}{\angstrom}$. It should be emphasized that our analysis applies to \textit{any} electronic systems, including disordered and/or interacting systems. This result demonstrates that our analysis is powerful in understanding quantum materials.  

Finally, for the sake of completeness, we discuss the relation of the quantum weight $K$ defined by static structure factor to quantum geometry. As first shown by SWM, the first-negative moment of optical conductivity $\Re W^1$ is related to the many-body quantum metric defined over the twisted boundary condition~\cite{souza_polarization_2000}. Then, based on the relation Eq.~\eqref{K-w1}, the quantum weight $K$ is also related to the many-body quantum metric. Here, we generalize the  work of SWM which considers systems with a unique ground state and introduce the many-body quantum metric $G$ for the general case of $r$-fold degenerated ground states (related to each other by spontaneous symmetry breaking or topological order): 
\begin{align}
    G_{\alpha\beta}(\vec{\theta}) &\equiv \frac{1}{r}\sum_{i=1}^{r}\Re\mel{\partial_{\alpha}\Psi_{i\vec{\theta}}}{(1-P_{\vec{\theta}})}{\partial_{\beta}\Psi_{i\vec{\theta}}}. 
\end{align}
Here, $\ket{\Psi_{i\theta}}$ is the $i$-th ground state under the twisted boundary condition specified by $\vec{\theta}$: $\Psi_{i\vec{\kappa}}(\vec{r}_1, \dots, \vec{r}_n + \vec{L}_{\mu}, \dots, \vec{r}_N) =  e^{i\theta_{\mu}}\Psi_{i\vec{\kappa}}(\vec{r}_1, \dots, \vec{r}_n, \dots, \vec{r}_N)$ where  $\vec{L}_{\mu}=(0,\dots, L_{\mu}, \dots,0)$ with $L_{\mu}$ the system size in $\mu$-direction, and $P_{\vec{\theta}}=\op{\Psi_{\vec{\theta}}}$ is the projector operator associated with the ground state. 

Assuming that the optical conductivity in the thermodynamic limit is insensitive to the choice of boundary condition $\theta$ or the particular ground state, one can show \cite{onishi_fundamental_2023} that the optical weight $\Re W^1$ and hence the quantum weight $K$ are related to $G$ as 
\begin{align}
    K &= 2\pi\int\frac{\dd^d{\vec{\theta}}}{(2\pi)^{d
    }} G(\vec{\theta}).
\end{align}

For noninteracting band insulators, the ground state is always unique and takes the form of a Slater determinant of occupied states over all wavevectors $\vec{k}$ in the Brillouin zone.  Then, 
the many-body quantum metric $G$ reduce to an integral in $k$-space: 
\begin{align}
    K_{\alpha\beta} &= 2\pi \int\frac{\dd^d{k}}{(2\pi)^d} g_{\alpha\beta}(\vec{k}).
\end{align}
where $g(\vec k)$ is a quantum metric in $k$-space defined as: 
\begin{align}
    g_{\alpha\beta}(\vec{k}) &= \Re[\mel{\partial_{\alpha}\Psi(\vec{k})}{(1-P(\vec{k}))}{\partial_{\beta}\Psi(\vec{k})}],
\end{align}
with $P(\vec{k})=\op{\Psi(\vec{k})}$ the projection operator onto the Slater determinant of occupied states at wavevector $\vec k$:  $\ket{\Psi(\vec{k})}=|u_1(\vec{k})\dots u_s(\vec{k})|$. Equivalently, $g(\vec k)$ is equal to the trace of non-Abelian quantum metric of occupied bands~\cite{peotta_superfluidity_2015}.  
This quantity appears as the gauge-invariant term in the localization functional of the Wannier functions~\cite{marzari_maximally_2012}. Thus, $K$ is related to the degree of localization of occupied electron states, consistent with our result on the quantum weight for strongly localized electron systems ($\hbar \rightarrow 0$) given above. We also note a relation between electron localization length and energy gap for noninteracting disordered systems in one dimension \cite{kivelson_wannier_1982}, which is a special case of our general relation between the quantum weight and the energy gap for all insulators.

To conclude, we have established quantum weight, defined through static structure factor, as a key material parameter that is connected to a variety of physical observables. The quantum weight represents the quantum fluctuation in electrons' center of mass. We derived its general relation to optical conductivity, dielectric constant, quantum geometry, and energy gap.  %
Our results apply to all insulators, including strongly correlated systems. Experimental determination by X-ray scattering as well as first-principles calculation of quantum weight for real materials~\cite{ghosh_probing_2024} are called for.

\begin{acknowledgements}
This work was supported by National Science Foundation (NSF) Convergence Accelerator
Award No. 2235945. YO is grateful for the support provided by the Funai Overseas Scholarship. 
LF was partly supported by the David and Lucile Packard Foundation. 
\end{acknowledgements}

\bibliography{references}

\newpage
\appendix
\begin{widetext}
\section{Supplemental Material}

\subsection{Quantum weight in atomic insulators}
To gain insight into the quantum weight, %
let us consider a solid made of an array of atoms %
that are far away from each other, so that the hopping between the atoms is negligible leading to trivial flat bands.  %
In such atomic insulators, optical absorption comes from electric dipole transitions between occupied and unoccupied energy levels within individual atoms. It is straightforward to show that the quantum weight $K$  %
is given by the ratio of the intra-atomic position fluctuation to the area of the unit cell $A_{0}$: 
\begin{align}
    K^{\rm AI}_{\alpha \beta} = & \frac{2\pi}{A_0} \mel{\Phi}{(r_{\alpha}-\expval{r_{\alpha}})(r_{\beta}-\expval{r_{\beta}})}{\Phi}  \label{K-AI}
\end{align}
where $\ket{\Phi}$ denotes the ground state of an atom, $\expval{r_\alpha}=\mel{\Phi}{r_{\alpha}}{\Phi}$ is the expectation value of $r_{\alpha}$ in the ground state, and $r_{\alpha} = \sum_{i}r_{\alpha, i}$ is the sum of the position of each electron. Notably, the right-hand side of Eq.~\eqref{K-AI} is precisely the polarization (=center-of-mass) fluctuation in an array of atoms, as discussed in the main text.

\subsection{Detailed derivation of the relation between the quantum weight and the optical weight}
Here we derive the following relation between the quantum weight $K$ and the negative-first moment of optical conductivity $W^1$ for general insulators:
\begin{align}
    \Re W^1 = \frac{e^2}{2\hbar} K, \label{eq:SM_W1K}
\end{align}
where $W^1$ is defined as 
\begin{align}
    W^1 = \int_0^{\infty} \dd{\omega} \frac{\sigma^{\rm abs}(\omega)}{\omega}
\end{align}
with the absorptive part of the conductivity tensor $\sigma^{\rm abs}(\omega) = (\sigma(\omega) + \sigma(\omega)^{\dagger})/2$
and the quantum weight is defined as the quadratic coefficient in $\vec{q}$ of the static structure factor $S_{\vec{q}}=(1/V)\expval{\rho_{\vec{q}}\rho_{-\vec{q}}}$ as 
\begin{align}
    S_{\vec{q}} &= \frac{e^2}{2\pi}K_{\alpha\beta} q_{\alpha}q_{\beta} + \dots.  
\end{align}
To show Eq.~\eqref{eq:SM_W1K}, we use the dissipation fluctuation theorem:
\begin{align}
    \Im \Pi(\vec{q}, \omega) &= S(\vec{q}, \omega)/(2\hbar), \label{eq:SM_DFT}
\end{align}
where $\Pi(\vec{q}, \omega)$ describes the density response to the external potential $V_{\rm ext}$, as defined in the main text. We will also use the following relation between $S(\vec{q},\omega)$ and $S(-\vec{q},-\omega)$:
\begin{align}
    S(\vec{q},\omega) = e^{\beta \hbar\omega} S(-\vec{q}, -\omega). \label{eq:Sw-w}
\end{align}
This relation can be shown as follows:
\begin{align}
    S(\vec{q},\omega) &= \frac{1}{V}\int_{-\infty}^{\infty}\dd{t} e^{i\omega t} \expval{\hat{\rho}_{\vec{q}}(t)\hat{\rho}_{-\vec{q}}(0)} \\
    &= \frac{1}{V}\sum_{n,m} \int_{-\infty}^{\infty}\dd{t} e^{i\omega t} \frac{e^{-\beta \ene_n}}{Z} \mel{n}{\hat{\rho}_{\vec{q}}(t)}{m}\mel{m}{\hat{\rho}_{-\vec{q}}(0)}{n} \\
    &= \frac{1}{V}\sum_{n,m} \int_{-\infty}^{\infty}\dd{t} e^{i(\omega + (\ene_n-\ene_m)/\hbar) t} \frac{e^{-\beta \ene_n}}{Z} \mel{n}{\hat{\rho}_{\vec{q}}}{m}\mel{m}{\hat{\rho}_{-\vec{q}}}{n} \\
    &= \frac{1}{V}\sum_{n,m} 2\pi \delta(\omega+\frac{\ene_n-\ene_m}{\hbar}) \frac{e^{-\beta \ene_n}}{Z} \mel{n}{\hat{\rho}_{\vec{q}}}{m}\mel{m}{\hat{\rho}_{-\vec{q}}}{n} \\
    &= \frac{1}{V}\sum_{n,m} 2\pi \delta(\omega-\frac{\ene_n-\ene_m}{\hbar}) \frac{e^{-\beta \ene_m}}{Z} \mel{m}{\hat{\rho}_{\vec{q}}}{n}\mel{n}{\hat{\rho}_{-\vec{q}}}{m} \\
    &= e^{\beta\hbar\omega}\frac{1}{V}\sum_{n,m} 2\pi \delta(-\omega+\frac{\ene_n-\ene_m}{\hbar}) \frac{e^{-\beta \ene_n}}{Z} \mel{n}{\hat{\rho}_{-\vec{q}}}{m}\mel{m}{\hat{\rho}_{\vec{q}}}{n} \\
    &= e^{\beta\hbar\omega} S(-\vec{q},-\omega).
\end{align}
Here, $\beta=(\kB T)^{-1}$ is the inverse temperature, $Z$ is the partition function, $\ket{n}$ is the $n$-th energy eigenstate with energy $\ene_n$.  
In particular, Eq.~\eqref{eq:Sw-w} implies $S(\vec{q},\omega) = 0$ for $\omega<0$ at zero temperature.

Due to the continuity equation $\pdv*{\rho}{t} + \nabla \cdot \vec{j} = 0$, $\Pi$ and $\sigma$ satisfies the following relation:
\begin{align}
    \Pi(\vec{q},\omega) &= i\frac{q_\alpha q_\beta \sigma_{\alpha\beta}(\vec{q},\omega)}{\omega}. \label{eq:SM_chi-sigma}
\end{align}
Noting that only the symmetric part of $\sigma$ contribute to $\Pi$, Eq.~\eqref{eq:SM_DFT} and Eq.~\eqref{eq:SM_chi-sigma} yield 
\begin{align}
    \frac{q_\alpha q_\beta \Re\sigma_{\alpha\beta}^{\rm abs}(\vec{q},\omega)}{\omega} = \frac{S(\vec{q},\omega) + S(\vec{q}, -\omega)}{2\hbar} \quad \text{for $\omega > 0$}
\end{align}
where we have used $S(\vec{q}, -\omega)=0$. Integrating over frequencies from $0$ to $\infty$, we will get
\begin{align}
    \int_0^{\infty}\dd{\omega}\frac{q_\alpha q_\beta \Re\sigma_{\alpha\beta}^{\rm abs}(\vec{q},\omega)}{\omega} = \int_{-\infty}^{\infty}\dd{\omega}\frac{S(\vec{q},\omega)}{2\hbar} = \frac{\pi}{\hbar}S_{\vec{q}}. 
\end{align}
Expanding in $\vec{q}$ around $\vec{q}=0$, we obtain 
\begin{align}
    q_\alpha q_\beta \Re W^1_{\alpha\beta} &= \frac{\pi}{\hbar} \frac{e^2}{2\pi} K_{\alpha\beta} q_{\alpha} q_{\beta},
\end{align}
which implies Eq.~\eqref{eq:SM_W1K}.

\end{widetext}

\end{document}